\documentclass[%
 aps,
 prapplied,
 a4paper,
 notitlepage,
 reprint,
 superscriptaddress,
 floatfix,
 raggedbottom,
]{revtex4-2}
\usepackage{amsmath,amssymb,amsfonts}

\usepackage{CJKutf8} 
\newcommand{\cname}[1]{\ {\rmfamily (}#1{\rmfamily )}}

\usepackage{subcaption}

\usepackage{xcolor}
\usepackage{url}

\usepackage{graphicx} 

\renewcommand{\eqref}[1]{Eq.~\ref{#1}}
\newcommand{\figref}[1]{Fig.~\ref{#1}}

\usepackage{dcolumn}
\usepackage{bm}

\usepackage{ragged2e}

\usepackage[colorlinks=true, 
            citecolor=blue, 
            linkcolor=red, 
            urlcolor=blue]{hyperref}

\begin{document}

\preprint{APS/}

\title{\textbf{Coupled Soft-Clamped Polygon Resonators with Tunable Mechanical Coupling} 
}%

\author{Zhihao Niu}
\affiliation{Quantum Science Center of Guangdong-Hong Kong-Macao Greater Bay Area, Shenzhen 518045, China}
\affiliation{Modern Matter Laboratory and Advanced Materials Thrust, The Hong Kong University of Science and Technology (Guangzhou), Guangzhou 511400, China}

\author{Peng Chen}
\affiliation{Quantum Science Center of Guangdong-Hong Kong-Macao Greater Bay Area, Shenzhen 518045, China}
\affiliation{Modern Matter Laboratory and Advanced Materials Thrust, The Hong Kong University of Science and Technology (Guangzhou), Guangzhou 511400, China}

\author{Shenggen Zheng}

\affiliation{Quantum Science Center of Guangdong-Hong Kong-Macao Greater Bay Area, Shenzhen 518045, China}

\author{Yuan-Yuan Zhao\begin{CJK}{UTF8}{gbsn}
\cname{赵圆圆}
\end{CJK}}
\email{zhaoyuanyuan1@quantumsc.cn}
\affiliation{Quantum Science Center of Guangdong-Hong Kong-Macao Greater Bay Area, Shenzhen 518045, China}

\author{Yuanyuan Zhao\begin{CJK}{UTF8}{gbsn}
\cname{赵媛媛}\end{CJK}}

\email{zhaoyuanyuan@quantumsc.cn}
\email{zhaoyuanyuan@gmail.com}
\affiliation{Quantum Science Center of Guangdong-Hong Kong-Macao Greater Bay Area, Shenzhen 518045, China}

\date{\today}

\begin{abstract}

Coupler geometry controls both the frequency splitting and the dissipation of coupled polygon resonators. We develop a continuum model for two identical square-shaped resonators coupled by a shared tether and compare the analytical predictions with finite element simulations. The model relates the coupler length and width to the frequency splitting and dissipation dilution of the two coupled modes. In the short-coupler limit, $k_cl_c\ll1$, the symmetric-mode coupler undergoes nearly uniform rotation, and its torsional strain energy vanishes asymptotically. The remaining loss contributions arise from polygon-side bending and supporting-tether torsion. The reduced supporting-tether contribution allows the symmetric-mode dissipation dilution to exceed that of an isolated polygon with the same side and supporting-tether dimensions. The antisymmetric mode retains torsional strain in the coupler and has an additional loss contribution. We also derive an expression for $Q\cdot f$ in the short-coupler and short-tether limits, where $Q$ is the mechanical quality factor and $f$ is the resonance frequency. This expression gives the dependence of $Q\cdot f$ on tensile stress, film thickness, and supporting-tether geometry. For a coupler length of $100~\mu\mathrm{m}$ and width of $2~\mu\mathrm{m}$, finite element calculations give a frequency splitting of $17.7~\mathrm{Hz}$ and a symmetric-mode dilution factor approximately $13\%$ above the analytical isolated-polygon value.

\end{abstract}
\maketitle

\section{Introduction}

The geometry of a mechanical coupler affects both the frequency splitting and the strain distribution of the coupled modes~\cite{catalini2020soft}. The strain distribution affects the mechanical loss and quality factor of each mode. Low dissipation is important for precision measurements \cite{mason2019continuous,xia2025motional,gonzalez1994brownian} and quantum control of mechanical motion \cite{wilson2015measurement,rossi2018measurement,ockeloen2018stabilized,palomaki2013entangling}. High $Q\cdot f$, where $Q$ is the mechanical quality factor and $f$ is the resonance frequency, is important for quantum-coherent operation \cite{aspelmeyer2014cavity,engelsen2024ultrahigh}.

Dissipation dilution raises the quality factors of strained resonators above the intrinsic material value \cite{fedorov2019generalized,unterreithmeier2010damping,schmid2011damping}. Static tension supplies restoring energy through the geometric extension induced by transverse motion \cite{engelsen2024ultrahigh}, while bending and torsional deformation contribute to material loss. Bending near rigid clamps can limit the resulting increase in $Q$. The contribution of boundary bending to mechanical loss can be reduced through phononic bandgap engineering \cite{tsaturyan2017ultracoherent,ghadimi2018elastic,beccari2022strained,cupertino2024centimeter}, clamp tapering \cite{bereyhi2019clamp}, hierarchical structuring \cite{fedorov2020fractal,bereyhi2022hierarchical,ciers2024nanomechanical}, and the use of polygon resonators \cite{bereyhi2022perimeter,niu2026predictive}.

Within the perimeter-mode approximation for polygon resonators, the out-of-plane displacement vanishes at the supporting-tether attachment points, while rotation at these points twists the tethers. Compact polygon resonators have reached room-temperature quality factors above $10^9$ \cite{bereyhi2022perimeter}. Integration with on-chip optical cavities has also enabled measurements of correlated frequency fluctuations and frequency-noise cancellation \cite{kharbanda2026chip}.

Coupled resonator systems have been investigated in several configurations. In phononic dimers, defect separation and orientation control the coupling strength~\cite{catalini2020soft}. In dual-mass torsional oscillators, the antisymmetric mode reduces sensitivity to environmental vibrations~\cite{manley2026dual}. For polygon resonators, a shared tether produces hybrid perimeter modes with different frequencies and torsional losses \cite{bereyhi2022perimeter}. Selecting the coupler dimensions requires the mode splitting and loss of both branches to be evaluated together. Comparison with an isolated polygon then shows whether the coupler increases or decreases the dissipation dilution.

In this work, we study two identical square-shaped polygon resonators connected by a shared tether. We develop a continuum model relating the coupler length and width to the frequency splitting and the bending and torsional contributions to dissipation dilution, and compare the analytical results with finite element simulations. In the short-coupler limit, the symmetric-mode coupler undergoes nearly uniform rotation and its torsional strain energy vanishes asymptotically. Reduced supporting-tether loss then allows the dissipation dilution to exceed that of an isolated polygon with the same side and supporting-tether dimensions. The antisymmetric mode instead twists the shared coupler and retains an additional torsional-loss contribution. We also derive an analytical expression for the symmetric-mode $Q\cdot f$ in terms of the polygon-side tensile stress, film thickness, and supporting-tether dimensions. The resulting expressions give the frequency splitting and dissipation dilution for a given set of coupler dimensions.

\section{Continuum model of coupled polygon resonators}
\label{continuum model}

\begin{figure}
 \centering
 \includegraphics[scale = 0.5]{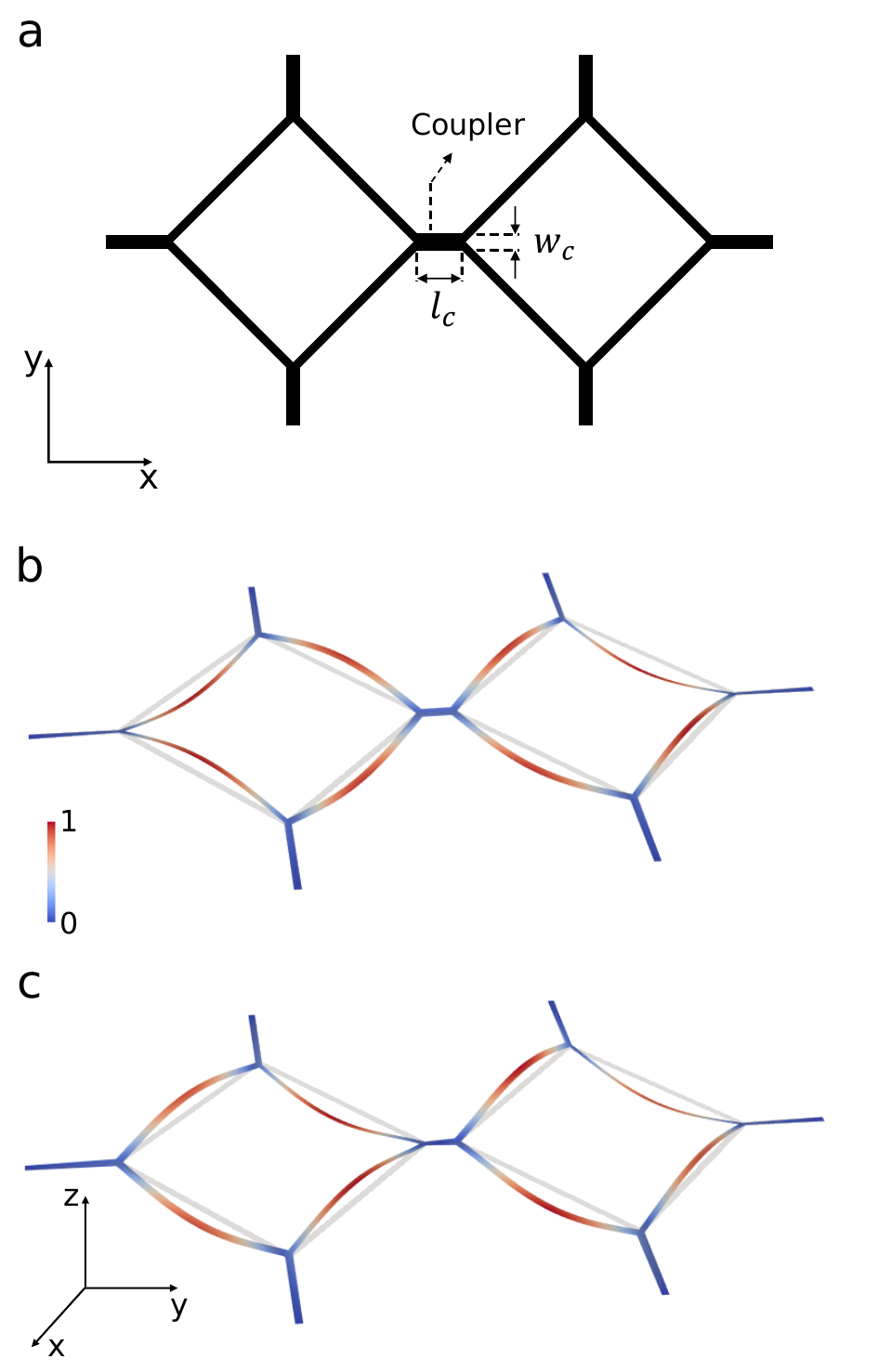}
 \caption{\justifying  Coupled perimeter modes in polygon resonators. 
(a) Schematic of two coupled polygon resonators connected by a shared tether. 
(b)(c) FEM-simulated displacement profiles of the symmetric and antisymmetric coupled modes, respectively.}

 \phantomsubcaption\label{fig1a} 
    \phantomsubcaption\label{fig1b}
    \phantomsubcaption\label{fig1c} 
    
 \label{fig1}
\end{figure}

We consider two identical square polygon resonators whose perimeter modes are coupled through torsion of a shared tether (\figref{fig1a}). In the high-tension limit, the transverse displacement $u_i(x_s)$ of polygon $i \in \{A,B\}$ along a side obeys

\begin{equation}
    -\sigma_s\frac{d^2u_i}{d x_s^2} =\rho\omega_0^2u_i,
\end{equation}

\noindent where $x_s$ is the coordinate along the polygon side, $\rho$ is the mass density, and $\omega_0$ is the uncoupled perimeter-mode frequency. The static axial stress $\sigma_s(x_s)\equiv\sigma_{x_sx_s}(x_s)$ is determined from force balance and the geometric constraint that the path length between the polygon clamping points remains fixed \cite{bereyhi2022perimeter}. For a slender side ($l_s \gg w_s, h$), we
approximate the mode shape by
\begin{equation}
    u_{i}(x_s) = u_{0,i} \phi_n(x_s),
    ~~\phi_n(x_s) = \sin{(\frac{n\pi x_s}{l_s})},
\end{equation}

\noindent where $u_{0,i}$ is the signed generalized modal amplitude, $n$ is the mode index, $l_s$ and $w_s$ are the side length and width, and $h$ is the film thickness.

The perimeter-mode motion twists each supporting tether. In the high-tension limit, the torsion angle $\tau_t$ of a supporting tether satisfies

\begin{equation}
    -(C_t + \sigma_{t}I_{y,t})\frac{d^2\tau_t}{dx_t^2} = \rho I_{p,t} \omega_0^2\tau_t,
\end{equation}

\noindent where $C_t$ is the intrinsic torsional rigidity, $I_{p,t}$ is the polar moment of inertia, $I_{y,t}$ is the second moment of area, $\sigma_t$ is the tensile stress in the supporting tether, and $x_t$ is the coordinate along the supporting tether. The torsion angle $\tau_t$ is subject to the boundary conditions
\begin{equation}
    \tau_t(0) = \sqrt{2}u_i^{\prime}(0),~~\tau_t(l_t)=0,
\end{equation}

\noindent with $l_t$ the tether length. A prime denotes differentiation
with respect to the relevant longitudinal coordinate.

The shared coupler has length $l_c$, width $w_c$, intrinsic
torsional rigidity $C_c$, axial stress $\sigma_c$, second moment $I_{y,c}$, and polar second moment $I_{p,c}$ (\figref{fig1a}). Its torsion angle $\tau_c(x_c)$ obeys the corresponding stress-modified wave equation

\begin{equation}
    -(C_c + \sigma_c I_{y,c})\frac{d^2 \tau_c}{d x_c^2} = \rho I_{p,c}\omega_0^2\tau_c.
    \label{eq:reduced TG equation}
\end{equation}

\noindent We define the effective torsional rigidity

\begin{equation}
    C_{\mathrm{eff}} = C_c + I_{y,c}\sigma_c,~~~ k_c = \omega_0 \sqrt{\frac{\rho I_{p,c}}{C_{\mathrm{eff}}}}.
    \label{C_eff}
\end{equation}

\noindent Here, $k_c$ is the torsional wavenumber. Because of the mirror symmetry of the structure, the coupled motion can be classified into symmetric (\figref{fig1b}) and antisymmetric modes (\figref{fig1c}). Transforming the coupled equations into the parity basis (see Appendix~\ref{Effecitive_dynamic_stiffness}) decouples the dynamics into two independent problems. The corresponding boundary conditions for the coupler become

\begin{equation}
    \tau_c(0) = \sqrt{2}u_A^{\prime}(0), ~~ \tau_c(l_c) = \sqrt{2}u_B^{\prime}(0).
    \label{coupler_boundary_condition}
\end{equation}
\noindent Under the local-coordinate convention used here, the endpoint rotations are equal for the symmetric mode ($u_{0,B}=u_{0,A}$) and opposite for the antisymmetric mode ($u_{0,B}=-u_{0,A}$).

\section{Frequency splitting of the coupled perimeter modes}

Using the continuum model established in Sec.~\ref{continuum model}, we obtain the coupler response from the endpoint torques associated with the torsional-wave solution (see Appendix~\ref{Effecitive_dynamic_stiffness}). For the common endpoint-rotation amplitude $\Theta$, the symmetric and antisymmetric dynamic coefficients are
\begin{equation}
    K_s^{\mathrm{dyn}} = -2C_{\mathrm{eff}}k_c\tan\frac{k_cl_c}{2},
\end{equation}
\noindent and
\begin{equation}
    K_a^{\mathrm{dyn}} = 2C_{\mathrm{eff}}k_c\cot\frac{k_cl_c}{2}.
\end{equation}
These coefficients include the coupler inertia and differ from the coefficients of its torsional strain energy. The wavenumber $k_c$ in these expressions is evaluated at the uncoupled frequency $\omega_0$ for the perturbative calculation below.

We assume that the coupling-induced frequency shifts are small compared with $\omega_0$. We also work away from torsional resonances of the coupler, so that its dynamic response varies smoothly over each shift. To first order, the coupled-mode frequencies are (see Appendix~\ref{subsec: frequency splitting})
\begin{equation}
    \omega_{s,a} \approx \omega_0 + \frac{n^2\pi^2K_{s,a}^{\mathrm{dyn}}}{2l_s^2M_p\omega_0},
    \label{coupled_fre}
\end{equation}
\noindent where $M_p$ is the effective modal mass of one uncoupled polygon resonator. The coupler inertia is already included in $K_{s,a}^{\mathrm{dyn}}$ and is not added separately to $M_p$. The frequency difference is therefore
\begin{equation}
    \Delta\omega=\omega_a-\omega_s
    \approx\frac{2n^2\pi^2C_{\mathrm{eff}}k_c}
    {l_s^2M_p\omega_0\sin(k_cl_c)}.
    \label{fre_splitting}
\end{equation}

In the short-coupler limit ($k_cl_c\ll1$), $K_s^{\mathrm{dyn}}\simeq-C_{\mathrm{eff}}k_c^2l_c$ and $K_a^{\mathrm{dyn}}\simeq4C_{\mathrm{eff}}/l_c-C_{\mathrm{eff}}k_c^2l_c/3$. The leading frequency splitting remains
\begin{equation}
    \Delta\omega \approx \frac{2n^2\pi^2}{l_s^2M_p\omega_0}\frac{C_{\mathrm{eff}}}{l_c}.
    \label{fre_splitting_sim}
\end{equation}
\noindent This approximation requires the accumulated torsional phase $k_cl_c$ to be small and the frequency shifts to remain perturbative relative to $\omega_0$.

For a rectangular coupler of fixed thickness, its intrinsic torsional rigidity scales as $C_c \approx Gw_ch^3/3$, where $G=E/[2(1+\nu)]$ is the shear modulus, $E$ is Young's modulus, and $\nu$ is Poisson's ratio. The following geometric scalings assume approximately unchanged polygon-side stress. If $C_c$ dominates $C_{\mathrm{eff}}$, \eqref{fre_splitting_sim} gives $\Delta\omega\propto w_c/l_c$. For a sufficiently wide coupler, the stress-induced geometric stiffness term $\sigma_cI_{y,c}$ can no longer be neglected. Using \eqref{stress_ribbon} and $I_{y,c}\propto hw_c^3$, we obtain $\sigma_cI_{y,c}\propto\sigma_sw_c^2$. When this term dominates the effective torsional rigidity, \eqref{fre_splitting_sim} predicts $\Delta\omega\propto w_c^2/l_c$, consistent with the scaling reported in \cite{bereyhi2022perimeter}.


\begin{figure*}[!tbp]
\centering
 \includegraphics[scale = 0.94]{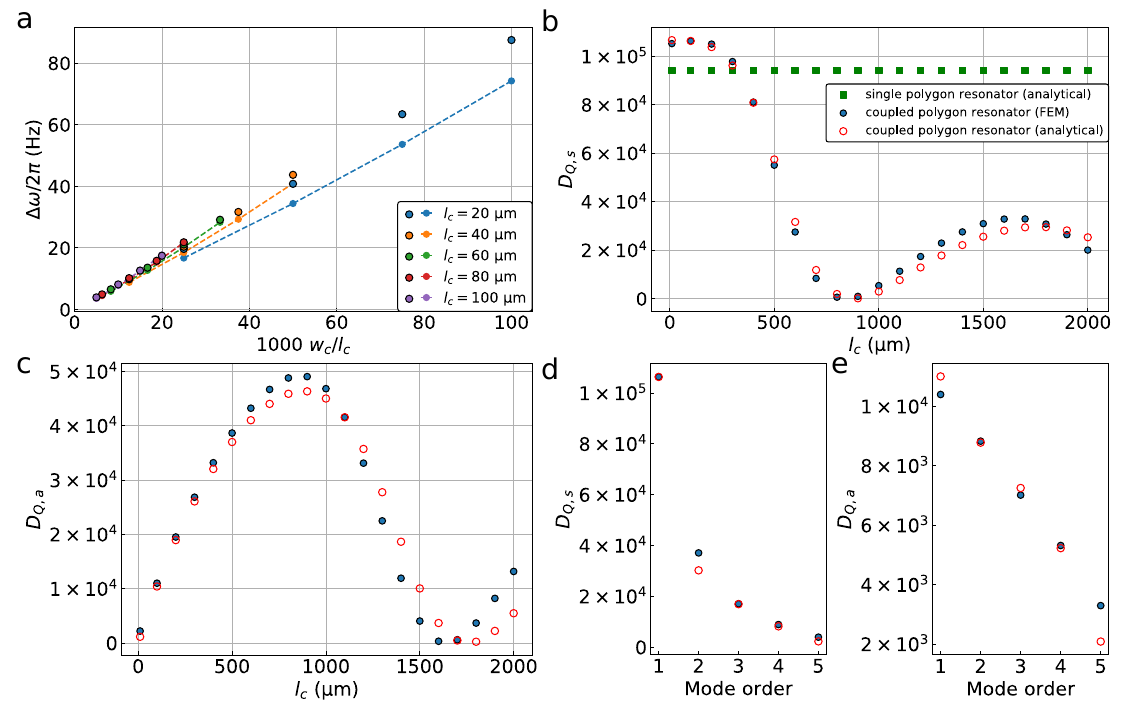}
 \caption{\justifying Comparison of analytical results and FEM simulations. 
(a) Frequency splitting between the symmetric and antisymmetric modes as a function of the scaled coupler width-to-length ratio $1000~w_c/l_c$ for different coupler lengths and widths. 
The blue, orange, green, red, and purple markers correspond to $l_c = 20$, 40, 60, 80, and $100~\mu\mathrm{m}$, respectively. Markers with black outlines indicate analytical predictions, whereas markers without outlines represent FEM results. 
(b)(c) $D_Q$ of the symmetric and antisymmetric modes, respectively, as a function of the coupler length $l_c$ at fixed $w_c=2~\mu\mathrm{m}$. 
(d)(e) Mode-order dependence of $D_Q$ for the symmetric and antisymmetric modes, respectively, at $l_c=100~\mu\mathrm{m}$ and $w_c=2~\mu\mathrm{m}$. 
In (b)--(e), solid blue circles denote FEM results, while open red circles indicate analytical predictions. The green squares in (b) represent the analytical results for an isolated polygon resonator whose polygon sides and supporting tethers have the same dimensions as those in the coupled system ($l_s=700~\mu\mathrm{m}$, $w_s=300~\mathrm{nm}$, $l_t=700~\mu\mathrm{m}$, and $w_t=400~\mathrm{nm}$).}
 \phantomsubcaption\label{fig2a} 
    \phantomsubcaption\label{fig2b}
    \phantomsubcaption\label{fig2c} 
    \phantomsubcaption\label{fig2d} 
    \phantomsubcaption\label{fig2e}    
 \label{fig2}
\end{figure*}

\section{Dissipation dilution factors}

Dissipation dilution enhances the quality factor $Q$ of highly strained mechanical resonators by increasing the fraction of elastic energy stored in tensile deformation, which does not contribute to mechanical loss at first order \cite{fedorov2019generalized}. The resulting quality factor is given by $Q=D_Q Q_{\mathrm{int}}$, where $Q_{\mathrm{int}}=1/\phi$ is the intrinsic quality factor associated with the material loss angle $\phi$ \cite{fedorov2019generalized}. The dissipation dilution factor $D_Q$ is defined as the ratio of the total stored elastic energy $\langle W_{\mathrm{total}}\rangle$ to the lossy elastic energy $\langle W_{\mathrm{lossy}}\rangle$ \cite{fedorov2019generalized}:
\begin{equation}
D_Q=\frac{\langle W_{\mathrm{total}}\rangle}{\langle W_{\mathrm{lossy}}\rangle}
=\frac{1}{\alpha\lambda+\beta\lambda^2},
\label{DQ_definition}
\end{equation}

\noindent where $\lambda = \frac{h}{l}\sqrt{\frac{E}{12\sigma}}$ is the strain parameter determined by resonator length $l$, thickness $h$, Young's modulus $E$, and static stress $\sigma$. The coefficients in the denominator $\alpha$ and $\beta$ are determined by the resonator's boundary curvature and the distributed bending over the rest of the resonator mode, respectively.

For square polygons ($N=4$), the continuum model yields analytical expressions for the $D_Q$ of the coupled perimeter modes. In the high-dilution limit, the total stored elastic energy is approximated by the tensile energy in the polygon sides. The lossy elastic energy comprises the distributed bending energy of the polygon sides and the torsional energies of the supporting tethers and the coupler. Evaluating these energy contributions gives the $D_Q$ of the symmetric and antisymmetric modes as (see Appendix~\ref{Appendix B})

\begin{equation}
\begin{split}
      D_{Q,s}^{-1} &= n^2\pi^2\lambda_l^2 + \lambda_l^2 \frac{3k_tl_{s}w_t}{2(1+\nu)w_s}\frac{2k_tl_{t}+\sin{(2k_tl_{t}})}{\sin^2(k_tl_{t})} \\
      &\quad + \lambda_l^2\frac{k_cl_sw_c}{2(1+\nu)w_s}\frac{k_cl_c-\sin{(k_cl_c)}}{\cos^2(k_cl_c/2)},
      \label{D_Qs}
\end{split}
\end{equation}

\noindent and

\begin{equation}
\begin{split}
    D_{Q,a}^{-1} &= n^2\pi^2\lambda_l^2 + \lambda_l^2 \frac{3k_tl_{s}w_t}{2(1+\nu)w_s}\frac{2k_tl_{t}+\sin{(2k_tl_{t}})}{\sin^2(k_tl_{t})} \\
      &\quad + \lambda_l^2\frac{k_cl_sw_c}{2(1+\nu)w_s}\frac{k_cl_c+\sin{(k_cl_c)}}{\sin^2(k_cl_c/2)},
      \label{D_Qa}
\end{split}
\end{equation}

\noindent where $k_t$ is the supporting-tether torsional wavenumber, $w_t$ is the supporting-tether width, $w_s$ is the polygon-side width, and $\lambda_l = \frac{h}{l_s}\sqrt{\frac{E}{12\sigma_s}}$. 

Retaining only the first term in each inverse dilution factor gives the upper bound $D_Q\leq(n^2\pi^2\lambda_l^2)^{-1}$; the remaining terms describe the torsional dissipation introduced by the supporting tethers and the coupler. At fixed polygon-side stress and modal shape, the supporting-tether contribution to $D_Q^{-1}$ is $25\%$ lower than in an isolated polygon. Although the coupler introduces additional torsional loss, its contribution to the symmetric mode vanishes asymptotically in the short-coupler limit.

In the short-coupler and short-tether limits, the symmetric-mode frequency approaches the uncoupled perimeter-mode frequency, $\omega_0=(n\pi/l_s)\sqrt{\sigma_s/\rho}$. Combining this result with \eqref{D_Qs}, and using $\lambda_l=(h/l_s)\sqrt{E/(12\sigma_s)}$, we obtain

\begin{equation}
    Q_sf_s\simeq
\frac{
6n l_s\sigma_s^{3/2}
}{
\phi E h^2\sqrt{\rho}
\left[
n^2\pi^2+
\displaystyle\frac{6l_sw_t}{(1+\nu)w_sl_t}
\right]
}.
    \label{Qf_short_limits}
\end{equation}

For fixed dimensionless loss terms and material loss angle, \eqref{Qf_short_limits} gives $Q\cdot f\propto\sigma_s^{3/2}/h^2$. At fixed polygon-side stress, the supporting-tether contribution can be further suppressed by reducing the ratio $w_t/l_t$ or increasing the polygon-side width $w_s$. Comparisons between different geometries should combine the stress relation in \eqref{eq:sigma_s} with \eqref{Qf_short_limits}.

Equations~(\ref{fre_splitting_sim}) and~(\ref{Qf_short_limits}) give the frequency splitting and symmetric-mode $Q\cdot f$ for the same geometry. The splitting depends on the effective torsional rigidity and length of the coupler, while the leading symmetric-mode loss contributions arise from polygon-side bending and torsion of the remaining supporting tethers. In this limit, changing the coupler dimensions varies the splitting while the coupler's torsional-loss contribution is suppressed. Changes in coupler geometry also redistribute the polygon-side stress and hence change $Q\cdot f$.

\section{Simulation results}

We compare analytical and FEM results for a high-stress stoichiometric silicon nitride (Si$_3$N$_4$) film with initial stress $\sigma_0 = 1~\mathrm{GPa}$, $E = 250~\mathrm{GPa}$, $\nu = 0.23$, $\rho = 3100~\mathrm{kg/m^3}$, and $h = 100~\mathrm{nm}$ at room temperature.

For each geometry, a static equilibrium calculation determines the redistributed stress field. This field is then used in a prestressed eigenfrequency analysis to obtain the mode frequencies and displacement fields. The dilution factor is evaluated according to \eqref{DQ_definition} as the ratio of the total modal elastic energy, including the prestress contribution, to the material strain energy associated with bending and torsion.

We first examine the frequency splitting between the symmetric and antisymmetric modes formed by coupling the fundamental modes of two identical polygon resonators. The polygon sides have length $l_s=700~\mu\mathrm{m}$ and width $w_s=300~\mathrm{nm}$, while the supporting tethers have length $l_t=700~\mu\mathrm{m}$ and width $w_t=400~\mathrm{nm}$. \figref{fig2a} compares the short-coupler prediction of \eqref{fre_splitting_sim} with FEM simulations for the frequency splitting as a function of the scaled width-to-length ratio, $1000~w_c/l_c$, for coupler widths $w_c=0.5$, 1, 1.5, and $2~\mu\mathrm{m}$, and coupler lengths $l_c = 20$, 40, 60, 80, and $100~\mu \mathrm{m}$. Both the intrinsic and stress-induced contributions to $C_{\mathrm{eff}}$ are retained. The analytical results show trends consistent with the FEM simulations over the investigated parameter range, with larger quantitative deviations for the shortest couplers ($l_c=20~\mu\mathrm{m}$). 

Figures~\ref{fig2b} and~\ref{fig2c} compare the analytical results with FEM simulations for the symmetric and antisymmetric modes, respectively, as a function of the coupler length $l_c$. The coupled resonator has $w_c=2~\mu\mathrm{m}$ and the same polygon-side and supporting-tether dimensions as the single-polygon reference, with $l_s=l_t=700~\mu\mathrm{m}$, $w_s=300~\mathrm{nm}$, and $w_t=400~\mathrm{nm}$. Both branches show a nonmonotonic length dependence, with a minimum in the symmetric-mode dilution and a maximum in the antisymmetric-mode dilution at intermediate lengths. The symmetric branch has the higher dilution for short couplers, while the ordering reverses over part of the longer-coupler range.

Figures~\ref{fig2a} and~\ref{fig2b} show the frequency splitting and symmetric-mode dilution factor for different coupler dimensions within the investigated range. For $w_c=2~\mu\mathrm{m}$ and $l_c=100~\mu\mathrm{m}$, the FEM results give a frequency splitting of approximately $17.7~\mathrm{Hz}$ and a symmetric-mode dilution factor of $1.06\times10^5$. The latter is about $13\%$ above the analytical value for an isolated polygon with the same side and supporting-tether dimensions. Increasing the coupler length to $500~\mu\mathrm{m}$ at the same width reduces the dilution factor to $5.5\times10^4$, below the isolated-polygon value. At a fixed length of $100~\mu\mathrm{m}$, increasing the width from $0.5$ to $2~\mu\mathrm{m}$ raises the analytical frequency splitting from approximately $3.9$ to $17.4~\mathrm{Hz}$; the corresponding FEM values increase from approximately $3.6$ to $17.7~\mathrm{Hz}$.

Figures~\ref{fig2d} and~\ref{fig2e} compare the analytical and FEM dilution factors as a function of the mode order $n$. The model reproduces the overall decrease in $D_Q$ with increasing mode order, but larger deviations occur for some higher-order modes, particularly the antisymmetric mode at $n=5$. These deviations limit the quantitative accuracy at higher mode orders; the short-coupler design comparisons above concern the fundamental modes.

\section{Conclusions and outlook}

We have calculated the frequency splitting and dissipation dilution of two polygon resonators connected by a shared tether. The continuum model includes the torsional restoring contributions from material elasticity and tensile stress, while material loss is evaluated from polygon-side bending and tether torsion. The analytical expressions and finite element calculations describe the dependence of frequency splitting on coupler dimensions and of dissipation dilution on coupler length and mode order.

The difference between the branches follows from their endpoint rotations. In the short-coupler limit, the symmetric-mode coupler undergoes nearly uniform rotation, and its torsional strain energy vanishes asymptotically. The external supporting tethers then provide the remaining torsional-loss contribution. At fixed stress and modal shape, replacing one fixed supporting tether of each polygon reduces this contribution to three quarters of the isolated-polygon value. This reduction can outweigh the additional coupler loss at finite length, producing the enhancement of symmetric-mode dissipation dilution shown in \figref{fig2b}. The antisymmetric mode twists the shared coupler and retains an additional loss contribution.

The coupler length therefore affects both the frequency splitting and the symmetric-mode dissipation dilution relative to an isolated polygon. In the short-coupler and short-tether limits, the analytical expression gives the dependence of $Q\cdot f$ on polygon-side tensile stress, film thickness, and the bending and supporting-tether loss terms. With these dimensionless loss terms and the material loss angle held fixed, $Q\cdot f$ scales as $\sigma_s^{3/2}/h^2$. Geometric changes also redistribute the static stress, and this dependence enters the comparison between different dimensions. A suitable coupler can therefore produce finite frequency splitting while retaining symmetric-mode dissipation dilution above the isolated-polygon value. The relation between coupling and dissipation may also be relevant to the design of mechanical networks \cite{wang2020collective,li2019honeycomblike} and soft-clamped phononic waveguides~\cite{xi2025soft}.

\section{Acknowledgments}

Y.-Y.~Z. is supported by the National Natural Science Foundation of China (Grant No.~12574401), the Guangdong Project (Grant No.~2024TQ08A680), and the Guangdong Provincial Quantum Science Strategic Initiative (Grant Nos.~GDZX2506001, GDZX2403002, GDZX2403001, GDZX2403005, GDZX2503001, and GDZX2503002). Y.~Z. acknowledges support from the National Natural Science Foundation of China (Grant Nos.~12474153 and 11804163) and the Guangdong Provincial Quantum Science Strategic Initiative (Grant No.~GDZX2401001).

\begin{appendix}

\makeatletter
\renewcommand{\thesubsection}{\thesection.\arabic{subsection}}
\renewcommand{\p@subsection}{}
\makeatother
\section{Derivation of the frequency splitting}
\label{derivation_frequency_splitting}

This appendix derives the analytical frequency splitting used in the main text. The torsional-wave solution gives the endpoint torques and hence the dynamic coefficients, including coupler inertia. These coefficients enter the modal frequency equations, which are expanded about the uncoupled polygon frequency.

\subsection{Stress-modified torsional dynamics of the coupler}
\label{continuum_model_coupler}

The coupler connecting the two identical $N$-sided polygon resonators is modeled as a stressed torsional beam with length $l_c$ and width $w_c$. Each polygon is suspended by $N-1$ supporting tethers of length $l_t$ and width $w_t$, while the remaining vertex connects to the coupler. In this structure, the polygons redistribute the tensile stress among the supporting branches. Force balance, together with the constraint that the path length between clamping points remains constant \cite{bereyhi2022perimeter}, yields the polygon-side stress
\begin{equation}
\sigma_s=\sigma_0(1-\nu)
\frac{1+2\bar r_l\sin(\pi/N)}
{1+4(w_s/l_s)\bar\gamma\sin^2(\pi/N)},
\label{eq:sigma_s}
\end{equation}
where $\bar r_l=[(N-1)l_t+l_c/2]/(Nl_s)$ is the branch-averaged length ratio and $\bar\gamma=[(N-1)l_t/w_t+l_c/(2w_c)]/N$ is the branch-averaged slenderness. The stress in the coupler can be expressed as

\begin{equation}
\sigma_c = 2\frac{w_s}{w_c}\sin(\pi/N)\sigma_s,
\label{stress_ribbon}
\end{equation}

\noindent The torsional deformation is governed by the stress-modified Timoshenko–Gere equation~\cite{niu2026predictive}
\begin{equation}
    EI_{\psi,c}\frac{d^4 \tau_c}{d x_c^4}+(\rho I_{\psi,c}\omega_0^2 - C_c-\sigma_cI_{y,c})\frac{d^2\tau_c}{d x_c^2}-\rho I_{p,c}\omega_0^2\tau_c = 0,
    \label{TG equation}
\end{equation}

\noindent where $I_{\psi,c}$ is the warping constant \cite{chopin2019extreme}. For the narrow coupler considered here, warping-related
corrections are neglected, giving \eqref{eq:reduced TG equation}. Its solution is

\begin{equation}
    \tau_c(x_c) = P\sin{(k_cx_c)} + Q\cos{(k_cx_c)},
\end{equation}

\noindent with $k_c$ defined in \eqref{C_eff}. Applying the endpoint rotations $\tau_c(0) = \Theta_A$ and $\tau_c(l_c) = \Theta_B$ gives

\begin{equation}
    \tau_c(x_c) = \Theta_A\cos{(k_cx_c)} + \frac{\Theta_B - \Theta_A\cos{(k_cl_c)}}{\sin{(k_cl_c)}}\sin{(k_cx_c)}.
    \label{coupler_torsion_angle}
\end{equation}

\noindent The endpoint rotations are related to the polygon modal
slopes by

\begin{equation}
    \Theta_i = \sqrt{2}u_i^{\prime}(0)=\sqrt{2}\frac{n\pi}{l_s}u_{0,i},~~~i = A, B.
\end{equation}

\noindent We use $\Theta=\Theta_A$ as the reference endpoint rotation; $\Theta_B$ follows the mode-parity convention specified after \eqref{coupler_boundary_condition}.

\subsection{Endpoint dynamic stiffness and material-loss energy}
\label{Effecitive_dynamic_stiffness}
To retain the frequency dependence of the endpoint response, we replace $\omega_0$ by a trial frequency $\omega$ in the torsional-wave equation and use $k_c(\omega)=\omega\sqrt{\rho I_{p,c}/C_{\mathrm{eff}}}$. The torques required to impose the endpoint rotations are
\begin{equation}
    T_A=-C_{\mathrm{eff}}\tau_c'(0),\qquad
    T_B=C_{\mathrm{eff}}\tau_c'(l_c).
    \label{coupler_endpoint_torques}
\end{equation}
Substitution of \eqref{coupler_torsion_angle} gives
\begin{equation}
    \begin{pmatrix}T_A\\T_B\end{pmatrix}
    =\frac{C_{\mathrm{eff}}k_c}{\sin(k_cl_c)}
    \begin{pmatrix}\cos(k_cl_c)&-1\\-1&\cos(k_cl_c)\end{pmatrix}
    \begin{pmatrix}\Theta_A\\\Theta_B\end{pmatrix}.
    \label{coupler_dynamic_matrix}
\end{equation}
With $\Theta_A=\Theta$, the generalized torques for equal and opposite endpoint rotations are, respectively,
\begin{equation}
    \begin{split}
        \mathcal{T}_s&=T_A+T_B=K_s^{\mathrm{dyn}}(\omega)\Theta,\\
        \mathcal{T}_a&=T_A-T_B=K_a^{\mathrm{dyn}}(\omega)\Theta.
    \end{split}
\end{equation}
For real harmonic amplitudes, define the potential and kinetic quadratic forms as $U_c=\tfrac12C_{\mathrm{eff}}\int_0^{l_c}(\tau_c')^2dx$ and $T_c=\tfrac12\rho I_{p,c}\omega^2\int_0^{l_c}\tau_c^2dx$. Integration by parts using the wave equation yields
\begin{equation}
    \begin{split}
        U_c-T_c&=\tfrac12(\Theta_AT_A+\Theta_BT_B)\\
        &=\tfrac12K_m^{\mathrm{dyn}}(\omega)\Theta^2,\qquad m=s,a.
    \end{split}
    \label{coupler_dynamic_energy}
\end{equation}
The dynamic coefficient therefore follows from $U_c-T_c$, rather than $U_c$ alone. Evaluation of the endpoint torques gives
\begin{equation}
    K_s^{\mathrm{dyn}}(\omega)=-2C_{\mathrm{eff}}k_c\tan\frac{k_cl_c}{2},
\end{equation}
\noindent and
\begin{equation}
    K_a^{\mathrm{dyn}}(\omega)=2C_{\mathrm{eff}}k_c\cot\frac{k_cl_c}{2}.
\end{equation}
The negative symmetric coefficient at small $k_cl_c$ is an inertial correction, not a negative material strain energy. Material loss is still evaluated from the positive torsional strain energy with the intrinsic rigidity $C_c$, as in Appendix~\ref{Appendix B}.

\subsection{Perturbative frequency splitting}
\label{subsec: frequency splitting}
Let $K_p$ and $M_p$ be the reference stiffness and modal mass of one uncoupled polygon, with $K_p=M_p\omega_0^2$. Since $\Theta=\sqrt{2}(n\pi/l_s)u_0$, the two modal frequency equations are
\begin{equation}
    2K_p-2M_p\omega_s^2
    +\frac{2n^2\pi^2}{l_s^2}K_s^{\mathrm{dyn}}(\omega_s)=0,
    \label{K_s}
\end{equation}
\noindent and
\begin{equation}
    2K_p-2M_p\omega_a^2
    +\frac{2n^2\pi^2}{l_s^2}K_a^{\mathrm{dyn}}(\omega_a)=0.
    \label{K_a}
\end{equation}
These equations include the distributed coupler inertia through the frequency-dependent coefficients. For small shifts and a smooth coupler response away from its torsional resonances, expansion about $\omega_0$ gives
\begin{equation}
    \omega_m\approx\omega_0+
    \frac{n^2\pi^2K_m^{\mathrm{dyn}}(\omega_0)}
    {2l_s^2M_p\omega_0},\qquad m=s,a.
\end{equation}
Using $K_a^{\mathrm{dyn}}-K_s^{\mathrm{dyn}}=4C_{\mathrm{eff}}k_c/\sin(k_cl_c)$ gives \eqref{fre_splitting}. Its leading short-coupler term is \eqref{fre_splitting_sim}. The inertial corrections to the endpoint coefficients scale as $l_c$, whereas the leading antisymmetric stiffness scales as $1/l_c$; their relative contribution to the leading splitting is of order $(k_cl_c)^2$.

\section{Energy decomposition for dissipation
dilution}
\label{Appendix B}

For the coupled square polygons ($N=4$) considered here, the dissipation dilution factor $D_Q$ is calculated using \eqref{DQ_definition} of the main text. The corresponding energies are evaluated by summing the contributions from the polygon sides, the supporting tethers, and the coupler. The lossless tensile energy is given by
\begin{equation}
    \langle W_{\mathrm{tens}}\rangle = 4\sigma_s w_sh\int_{0}^{l_s}[u^{\prime}(x)]^2dx.
\end{equation}

\noindent The lossy energy consists of the distributed bending energy of the polygon sides, and the torsional energy in the supporting tethers and the coupler:
\begin{equation}
    \langle W_{\mathrm{lossy}}\rangle = \langle W_{\mathrm{bend}}\rangle + \langle W_{\mathrm{tor},t} \rangle + \langle W_{\mathrm{tor},c} \rangle .
\end{equation}

\noindent The distributed bending energy is 
\begin{equation}
    \langle W_{\mathrm{bend}}\rangle = \frac{1}{3}Ew_sh^3\int_{0}^{l_s}[u^{\prime\prime}(x)]^2dx.
\end{equation}

\noindent The torsional contribution of the supporting tethers is
\begin{equation}
    \langle W_{\mathrm{tor},t}\rangle = \frac{Ew_th^3}{2(1+\nu)}\int_{0}^{l_t}[\tau_t^{\prime}(x)]^2dx,
\end{equation}

\noindent The torsional energy of the coupler is 

\begin{equation}
    \langle W_{\mathrm{tor},c} \rangle = \frac{1}{2}C_{c}\int_{0}^{l_c}[\tau_c^{\prime}(x)]^2dx
    \label{torsional_energy_coupler_loss}
\end{equation}

Collecting these terms yields a dissipation dilution factor of 
\begin{equation}
    D_{Q}^{-1} = D_{Q,\mathrm{bend}}^{-1} + D_{Q,\mathrm{tor},t}^{-1} +D_{Q,\mathrm{tor},c}^{-1},
\end{equation}

\noindent where 
\begin{equation}
    D_{Q,\mathrm{bend}}^{-1} = n^2\pi^2\lambda_l^2,
\end{equation}
\begin{equation}
    D_{Q,\mathrm{tor},t}^{-1} =  \lambda_l^2 \frac{3k_tl_{s}w_t}{2(1+\nu)w_s}\frac{2k_tl_{t}+\sin{(2k_tl_{t}})}{\sin^2(k_tl_{t})},
\end{equation}

\noindent and 
\begin{equation}
    D_{Q,\mathrm{tor},c}^{-1} = \lambda_l^2\frac{k_cl_sw_c}{2(1+\nu)w_s}\frac{k_cl_c-\sin{(k_cl_c)}}{\cos^2(k_cl_c/2)},
\end{equation}

\noindent for the symmetric mode and 
\begin{equation}
    D_{Q,\mathrm{tor},c}^{-1} =\lambda_l^2\frac{k_cl_sw_c}{2(1+\nu)w_s}\frac{k_cl_c+\sin{(k_cl_c)}}{\sin^2(k_cl_c/2)},
\end{equation}

\noindent for the antisymmetric mode. In the symmetric short-coupler limit, $k_cl_c - \sin(k_cl_c) \simeq (k_cl_c)^3/6$, so the coupler-loss term vanishes as $l_c \to 0$ at
fixed $k_c$ and endpoint rotation. This asymptotic suppression concerns the coupler contribution only; the bending
and supporting-tether contributions remain finite.

\end{appendix}

\bibliography{references-citation-revised}
\end{document}